\documentclass[10pt, conference, letterpaper]{IEEEtran}
\IEEEoverridecommandlockouts

\usepackage{fancyhdr}
\usepackage{epsfig}
\usepackage{threeparttable}
\usepackage{epsf,epsfig}
\usepackage{amsthm}
\usepackage[cmex10]{amsmath}
\usepackage{amssymb, amsfonts}
\usepackage[noadjust]{cite}
\usepackage{dsfont}
\usepackage{subfigure}
\usepackage{color}
\usepackage{soul}
\usepackage{amssymb}
 \usepackage{accents}
  \usepackage{algorithm}
   \usepackage[english]{babel}
    \usepackage{url}
    \usepackage{framed} 
        \usepackage{bm} 

\allowdisplaybreaks[4]

\DeclareMathSizes{10}{9.5}{5}{3}

\begin{document}
\newtheorem{theorem}{Theorem}
\newtheorem{acknowledgement}[theorem]{Acknowledgement}
\newtheorem{axiom}[theorem]{Axiom}
\newtheorem{case}[theorem]{Case}
\newtheorem{claim}[theorem]{Claim}
\newtheorem{conclusion}[theorem]{Conclusion}
\newtheorem{condition}[theorem]{Condition}
\newtheorem{conjecture}[theorem]{Conjecture}
\newtheorem{criterion}[theorem]{Criterion}
\newtheorem{definition}{Definition}
\newtheorem{exercise}[theorem]{Exercise}
\newtheorem{lemma}{Lemma}
\newtheorem{corollary}{Corollary}
\newtheorem{notation}[theorem]{Notation}
\newtheorem{problem}[theorem]{Problem}
\newtheorem{proposition}{Proposition}
\newtheorem{scheme}{Scheme}   %%% DZ added
\newtheorem{solution}[theorem]{Solution}
\newtheorem{summary}[theorem]{Summary}
\newtheorem{assumption}{Assumption}
\newtheorem{example}{\bf Example}
\newtheorem{remark}{\bf Remark}

\def\qed{$\Box$}
\def\QED{\mbox{\phantom{m}}\nolinebreak\hfill$\,\Box$}
\def\proof{\noindent{\emph{Proof:} }}
\def\poof{\noindent{\emph{Sketch of Proof:} }}
\def
\endproof{\hspace*{\fill}~\qed
\par
\endtrivlist\unskip}
\def\endproof{\hspace*{\fill}~\qed\par\endtrivlist\vskip3pt}

\def\E{\mathsf{E}}
\def\eps{\varepsilon}
\def\phi{\varphi}
\def\Lsp{{\boldsymbol L}}
\def\Bsp{{\boldsymbol B}}
\def\lsp{{\boldsymbol\ell}}
\def\Ltsp{{\Lsp^2}}
\def\Lpsp{{\Lsp^p}}
\def\Linsp{{\Lsp^{\infty}}}
\def\LtR{{\Lsp^2(\Rst)}}
\def\ltZ{{\lsp^2(\Zst)}}
\def\ltsp{{\lsp^2}}
\def\ltZt{{\lsp^2(\Zst^{2})}}
\def\ninN{{n{\in}\Nst}}
\def\oh{{\frac{1}{2}}}
\def\grass{{\cal G}}
\def\ord{{\cal O}}
\def\dist{{d_G}}
\def\conj#1{{\overline#1}}
\def\ntoinf{{n \rightarrow \infty }}
\def\toinf{{\rightarrow \infty }}
\def\tozero{{\rightarrow 0 }}
\def\trace{{\operatorname{trace}}}
\def\ord{{\cal O}}
\def\UU{{\cal U}}
\def\rank{{\operatorname{rank}}}
\def\acos{{\operatorname{acos}}}

\def\SINR{\mathsf{SINR}}
\def\SNR{\mathsf{SNR}}
\def\SIR{\mathsf{SIR}}
\def\tSIR{\widetilde{\mathsf{SIR}}}
\def\Ei{\mathsf{Ei}}
\def\l{\left}
\def\r{\right}
\def\({\left(}
\def\){\right)}
\def\lb{\left\{}
\def\rb{\right\}}

\setcounter{page}{1}

% Definitions
\newcommand{\eref}[1]{(\ref{#1})}
\newcommand{\fig}[1]{Fig.\ \ref{#1}}

% Bold lowercase
\def\bydef{:=}
\def\ba{{\mathbf{a}}}
\def\bb{{\mathbf{b}}}
\def\bc{{\mathbf{c}}}
\def\bd{{\mathbf{d}}}
\def\bee{{\mathbf{e}}}
\def\bff{{\mathbf{f}}}
\def\bg{{\mathbf{g}}}
\def\bh{{\mathbf{h}}}
\def\bi{{\mathbf{i}}}
\def\bj{{\mathbf{j}}}
\def\bk{{\mathbf{k}}}
\def\bl{{\mathbf{l}}}
\def\bn{{\mathbf{n}}}
\def\bo{{\mathbf{o}}}
\def\bp{{\mathbf{p}}}
\def\bq{{\mathbf{q}}}
\def\br{{\mathbf{r}}}
\def\bs{{\mathbf{s}}}
\def\bt{{\mathbf{t}}}
\def\bu{{\mathbf{u}}}
\def\bv{{\mathbf{v}}}
\def\bw{{\mathbf{w}}}
\def\bx{{\mathbf{x}}}
\def\by{{\mathbf{y}}}
\def\bz{{\mathbf{z}}}
\def\b0{{\mathbf{0}}}

% Bold capital letters
\def\bA{{\mathbf{A}}}
\def\bB{{\mathbf{B}}}
\def\bC{{\mathbf{C}}}
\def\bD{{\mathbf{D}}}
\def\bE{{\mathbf{E}}}
\def\bF{{\mathbf{F}}}
\def\bG{{\mathbf{G}}}
\def\bH{{\mathbf{H}}}
\def\bI{{\mathbf{I}}}
\def\bJ{{\mathbf{J}}}
\def\bK{{\mathbf{K}}}
\def\bL{{\mathbf{L}}}
\def\bM{{\mathbf{M}}}
\def\bN{{\mathbf{N}}}
\def\bO{{\mathbf{O}}}
\def\bP{{\mathbf{P}}}
\def\bQ{{\mathbf{Q}}}
\def\bR{{\mathbf{R}}}
\def\bS{{\mathbf{S}}}
\def\bT{{\mathbf{T}}}
\def\bU{{\mathbf{U}}}
\def\bV{{\mathbf{V}}}
\def\bW{{\mathbf{W}}}
\def\bX{{\mathbf{X}}}
\def\bY{{\mathbf{Y}}}
\def\bZ{{\mathbf{Z}}}

% mathbb Bold capital letters
\def\mA{{\mathbb{A}}}
\def\mB{{\mathbb{B}}}
\def\mC{{\mathbb{C}}}
\def\mD{{\mathbb{D}}}
\def\mE{{\mathbb{E}}}
\def\mF{{\mathbb{F}}}
\def\mG{{\mathbb{G}}}
\def\mH{{\mathbb{H}}}
\def\mI{{\mathbb{I}}}
\def\mJ{{\mathbb{J}}}
\def\mK{{\mathbb{K}}}
\def\mL{{\mathbb{L}}}
\def\mM{{\mathbb{M}}}
\def\mN{{\mathbb{N}}}
\def\mO{{\mathbb{O}}}
\def\mP{{\mathbb{P}}}
\def\mQ{{\mathbb{Q}}}
\def\mR{{\mathbb{R}}}
\def\mS{{\mathbb{S}}}
\def\mT{{\mathbb{T}}}
\def\mU{{\mathbb{U}}}
\def\mV{{\mathbb{V}}}
\def\mW{{\mathbb{W}}}
\def\mX{{\mathbb{X}}}
\def\mY{{\mathbb{Y}}}
\def\mZ{{\mathbb{Z}}}

% Caligraphic capital letters
\def\cA{\mathcal{A}}
\def\cB{\mathcal{B}}
\def\cC{\mathcal{C}}
\def\cD{\mathcal{D}}
\def\cE{\mathcal{E}}
\def\cF{\mathcal{F}}
\def\cG{\mathcal{G}}
\def\cH{\mathcal{H}}
\def\cI{\mathcal{I}}
\def\cJ{\mathcal{J}}
\def\cK{\mathcal{K}}
\def\cL{\mathcal{L}}
\def\cM{\mathcal{M}}
\def\cN{\mathcal{N}}
\def\cO{\mathcal{O}}
\def\cP{\mathcal{P}}
\def\cQ{\mathcal{Q}}
\def\cR{\mathcal{R}}
\def\cS{\mathcal{S}}
\def\cT{\mathcal{T}}
\def\cU{\mathcal{U}}
\def\cV{\mathcal{V}}
\def\cW{\mathcal{W}}
\def\cX{\mathcal{X}}
\def\cY{\mathcal{Y}}
\def\cZ{\mathcal{Z}}
\def\cd{\mathcal{d}}
\def\Mt{M_{t}}
\def\Mr{M_{r}}
%% my defs
\def\O{\Omega_{M_{t}}}
\newcommand{\figref}[1]{{Fig.}~\ref{#1}}
\newcommand{\tabref}[1]{{Table}~\ref{#1}}

%% From Kaibin
\newcommand{\var}{\mathsf{var}}
\newcommand{\fb}{\tx{fb}}
\newcommand{\nf}{\tx{nf}}
\newcommand{\BC}{\tx{(bc)}}
\newcommand{\MAC}{\tx{(mac)}}
\newcommand{\Pout}{p_{\mathsf{out}}}
\newcommand{\nnn}{\nn\\}
\newcommand{\FB}{\tx{FB}}
\newcommand{\TX}{\tx{TX}}
\newcommand{\RX}{\tx{RX}}
\renewcommand{\mod}{\tx{mod}}
\newcommand{\m}[1]{\mathbf{#1}}
\newcommand{\td}[1]{\tilde{#1}}
\newcommand{\sbf}[1]{\scriptsize{\textbf{#1}}}
\newcommand{\stxt}[1]{\scriptsize{\textrm{#1}}}
\newcommand{\suml}[2]{\sum\limits_{#1}^{#2}}
\newcommand{\sumlk}{\sum\limits_{k=0}^{K-1}}
\newcommand{\eqhsp}{\hspace{10 pt}}
\newcommand{\tx}[1]{\texttt{#1}}
\newcommand{\Hz}{\ \tx{Hz}}
\newcommand{\sinc}{\tx{sinc}}
\newcommand{\tr}{\mathrm{tr}}
\newcommand{\diag}{\mathrm{diag}}
\newcommand{\MAI}{\tx{MAI}}
\newcommand{\ISI}{\tx{ISI}}
\newcommand{\IBI}{\tx{IBI}}
\newcommand{\CN}{\tx{CN}}
\newcommand{\CP}{\tx{CP}}
\newcommand{\ZP}{\tx{ZP}}
\newcommand{\ZF}{\tx{ZF}}
\newcommand{\SP}{\tx{SP}}
\newcommand{\MMSE}{\tx{MMSE}}
\newcommand{\MINF}{\tx{MINF}}
\newcommand{\RC}{\tx{MP}}
\newcommand{\MBER}{\tx{MBER}}
\newcommand{\MSNR}{\tx{MSNR}}
\newcommand{\MCAP}{\tx{MCAP}}
\newcommand{\vol}{\tx{vol}}
\newcommand{\ah}{\hat{g}}
\newcommand{\tg}{\tilde{g}}
\newcommand{\teta}{\tilde{\eta}}
\newcommand{\heta}{\hat{\eta}}
\newcommand{\uh}{\m{\hat{s}}}
\newcommand{\eh}{\m{\hat{\eta}}}
\newcommand{\hv}{\m{h}}
\newcommand{\hh}{\m{\hat{h}}}
\newcommand{\Po}{P_{\mathrm{out}}}
\newcommand{\Poh}{\hat{P}_{\mathrm{out}}}
\newcommand{\Ph}{\hat{\gamma}}
\newcommand{\mat}[1]{\begin{matrix}#1\end{matrix}}
\newcommand{\ud}{^{\dagger}}
\newcommand{\C}{\mathcal{C}}
\newcommand{\nn}{\nonumber}
\newcommand{\nInf}{U\rightarrow \infty}

\addtolength{\textfloatsep}{-5mm}
\setlength{\abovecaptionskip}{-0.8pt}
\setlength{\belowcaptionskip}{-18pt}
\textheight=23.9cm
\voffset 1mm

%% linespace of equations
\setlength\abovedisplayskip{1.5 pt}
 \setlength\belowdisplayskip{1.5 pt}

%\DeclareMathSizes{10}{9.5}{5}{3}

%
%\title{\huge Wireless Data Acquisition for Edge Learning: \\
%Importance Aware Retransmission 
%\vspace{-10pt}
%} 
%\author{Dongzhu Liu, Guangxu Zhu, Jun Zhang and Kaibin Huang
%\thanks{\noindent  D. Liu, G. Zhu and K. Huang are with the Department of Electrical and Electronic Engineering, The University of Hong Kong, Pok Fu Lam, Hong Kong (Email: dzliu@eee.hku.hk, gxzhu@eee.hku.hk, haungkb@eee.hku.hk).  On going draft.
%}}
%\vspace{-10pt}}
%
%\maketitle

\title{\huge Leveraging Channel Noise for Sampling and Privacy via Quantized Federated Langevin Monte Carlo  }

%\author{Dongzhu Liu, Guangxu Zhu, Jun Zhang and Kaibin Huang
%\thanks{\noindent  D. Liu, G. Zhu and K. Huang are with the Dept. of Electrical and Electronic Engineering at the University of Hong Kong, Hong Kong. Email: \{dzliu,gxzhu,huangkb\}@eee.hku.hk.  
%
%J. Zhang is with the Dept. of Electronic and Computer Engineering at the Hong Kong University of Science and Technology, Hong Kong. Email: eejzhang@ust.hk. 
%
%}}

\author{ \IEEEauthorblockN{Yunchuan Zhang\IEEEauthorrefmark{1}, Dongzhu Liu\IEEEauthorrefmark{2}, and Osvaldo Simeone\IEEEauthorrefmark{1} \thanks{This work was supported by the European Research Council (ERC) under the European Union's Horizon 2020 Research and Innovation Programme (grant agreement No. 725732). The work of Yunchuan Zhang was supported by the China Scholarship Council and King's College London for their Joint Full-Scholarship (K-CSC) under Grant CSC202008420204.}}
\IEEEauthorblockA{\IEEEauthorrefmark{1} Department of Engineering, King's College London\\
\IEEEauthorrefmark{2}{School of Computing Science, University of Glasgow}\\
\{yunchuan.zhang, osvaldo.simeone\}@kcl.ac.uk, dongzhu.liu@glasgow.ac.uk.}
}
\maketitle

 \thispagestyle{fancy} % IEEE???\maketitle??????\thispagestyle{plain}?
                            % ???????????????plain???fancy
%      \lhead{} % ????????????{}???
%      \chead{} % ???
%      \rhead{} % ???
%      \lfoot{} % ???
%      \cfoot{} % ???
      \cfoot{\thepage} %????\thepage ??????
      \renewcommand{\headrulewidth}{0pt} %??0pt???????????
      \renewcommand{\footrulewidth}{0pt} %??0pt???????????
      
          \pagestyle{fancy}
      \cfoot{\thepage}

\begin{abstract}
For engineering applications of artificial intelligence, Bayesian learning holds significant advantages over standard frequentist learning, including the capacity to quantify uncertainty. Langevin Monte Carlo (LMC) is an efficient gradient-based approximate Bayesian learning strategy that aims at producing samples drawn from the posterior distribution of the model parameters. Prior work focused on a distributed implementation of LMC over a multi-access wireless channel via analog modulation. In contrast, this paper proposes quantized federated LMC (FLMC), which integrates  one-bit stochastic quantization of the local gradients with  channel-driven sampling. Channel-driven sampling leverages channel noise for the purpose of contributing to Monte Carlo sampling, while also serving the role of privacy mechanism. Analog and digital implementations of wireless LMC are compared as a function of differential privacy (DP) requirements, revealing the advantages of the latter at sufficiently high signal-to-noise ratio.
\end{abstract}

\begin{IEEEkeywords}
    Federated learning, Differential privacy, Langevin Monte Carlo, Power allocation
\end{IEEEkeywords}

\section{Introduction}

\begin{figure*}[t]

  \centering
  \centerline{\includegraphics[scale=0.22]{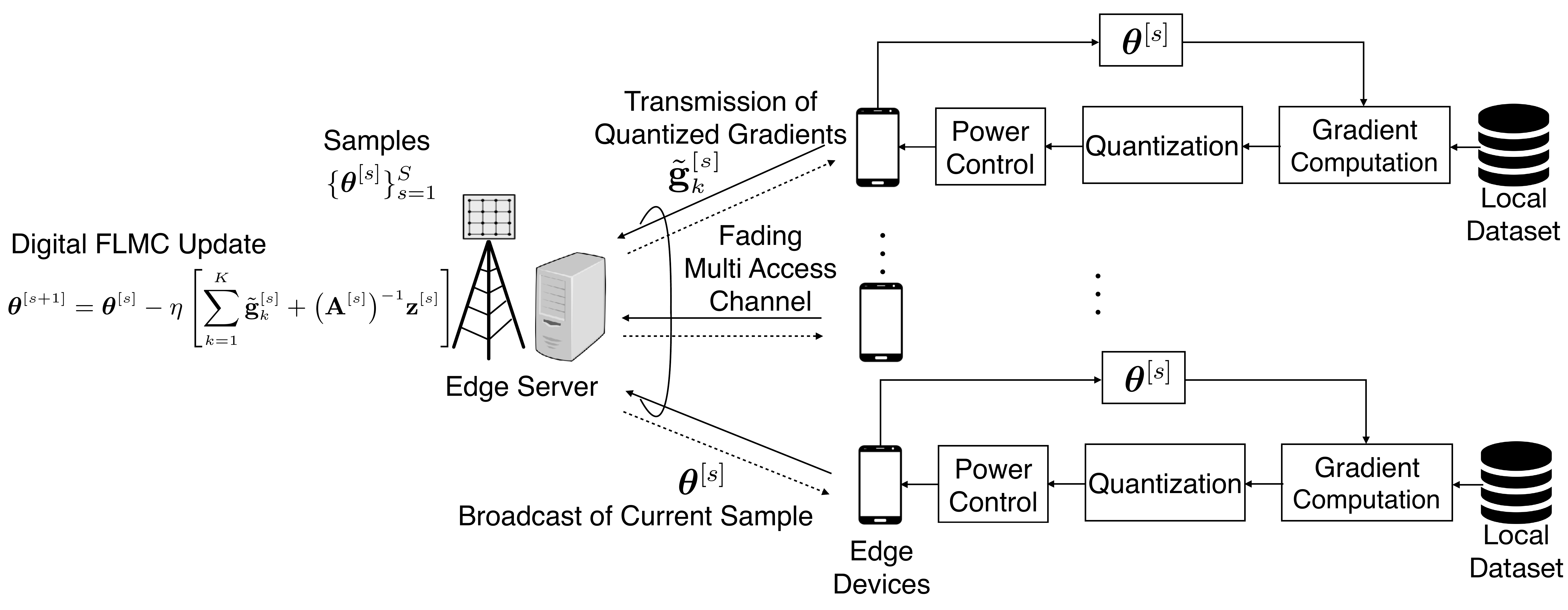}}
  \vspace{-0.0cm}
  \caption{Differentially private quantized federated Bayesian learning system based on LMC.}\label{fig: system model}
%
%\vspace{-0.2cm}
\end{figure*}

Federated learning (FL) is a distributed learning paradigm whereby  multiple devices  coordinate to  train a target global model, while avoiding the direct sharing of local data with the cloud  \cite{park2019wireless, zhou2019edge, zhu2020toward}. Prior work on wireless FL mainly focuses on conventional \emph{frequentist} learning, which produces point estimates of model parameter vectors by minimizing empirical loss metrics \cite{sery2021over, yang2021revisiting, amiri2020machine, liu2020privacy, cao2020optimized,zhu2020one}. In many engineering applications characterized by the availability of limited data and by the need to quantify uncertainty, \emph{Bayesian} learning provides a more effective and principled framework to define the learning problem (see, e.g., \cite{khan2021bayesian}).  Bayesian learning assigns a probability distribution to the model parameters, rather than collapsing any residual uncertainty in the model parameter space to a single point estimate. In this paper, we focus on the distributed implementation of Bayesian learning  in wireless systems within a federated learning setting, with the main goal of leveraging the wireless channel as part of the ``compute continuum'' between devices and server \cite{alwasel2021iotsim} (see Fig \ref{fig: system model}).

% Bayesian learning is widely implemented in wireless communications such as variational inference used for channel estimation \cite{wang2020vbi}, LMC via multiple access transmissions \cite{liu2021wireless}, a state-of-the-art gradient-based Markov chain Monte Carlo (MCMC) sampling approach that imposes Gaussian noise to gradient descent process. Here, we will focus on the comparison between analog FLMC and proposed quantized FLMC in wireless communications.

Scalable Bayesian learning solutions are either based on variational inference, whereby the distribution over the model parameters is optimized by minimizing a free energy metric \cite{jose2021free}; or on Monte Carlo (MC) sampling, whereby the distribution over the model parameters is represented by random samples \cite{angelino2016patterns}. It was recently pointed out in \cite{liu2021wireless} that MC solutions enable a novel interpretation of the wireless channel as part of the MC sampling process. In particular, reference \cite{liu2021wireless} proposed a Bayesian federated learning protocol based on Langevin MC (LMC), a noise-perturbed gradient-based MC strategy \cite{angelino2016patterns}, and \emph{analog} transmission. The paper demonstrated the role of the channel noise as a contributor to the LMC update, as well as a privacy mechanism (see also \cite{koda2020differentially,liu2021privacy}). In this paper, we devise an alternative strategy that implements LMC in a federated setting via \emph{digital} modulation under privacy constraints.

% Although each edge device only shares model updates instead of raw data in FL, it is still not sufficient to ensure privacy for local datasets. Differential privacy in uncoded transmission, which randomizes the datasets by introducing wireless channel noise was proved to serve as a privacy-preserving mechanism . Furthermore, the wireless channel noise was also shown to act as seed randomness in wireless FLMC which refers to channel-driven sampling method in \cite{liu2021wireless}.

% \subsection{Related Work}

Federated learning has been widely studied for implementation on wireless channels (see, e.g., \cite{gunduz2019machine}). Techniques that leverage the wireless channel for computation include over-the-air computation (AirComp), whereby superposition in non-orthogonal multiple access (NOMA) is used as a means to aggregate information from different sources \cite{nazer2007computation, cao2020optimized, liu2020aircomp}; channel noise for privacy, which enforces differential privacy (DP)  guarantees via power control \cite{seif2020wireless,liu2021privacy}; and channel noise for sampling, which was introduced above \cite{liu2021wireless}. Also related to this work are DP mechanisms based on stochastic quantization  \cite{gandikota2019vqsgd}.

% Several recent works provide analysis on wireless federated learning with DP guarantee. To further reduce the communication load, model compression techniques, mainly sparsification and quantization methods, are combined with Aircomp in \cite{alistarh2017qsgd, stich2018}.In this work Due to the fact that malicious server can potentially infer sensitive training datasets via membership inference attacks \cite{shokri2017membership} and model inversion attack \cite{fredrikson2015model}, higher privacy guarantee is achieved in \cite{Mohamed2021privacy} which boosts the DP with user sampling policy based on channel conditions. Moreover,

%  To the best of our knowledge, work in \cite{liu2020privacy} is the first to formally introduce analog wireless FLMC with DP analysis as well as adaptive power allocation policy, and to reproduce channel noise for privacy-preserving mechanism and efficient gradient aggregation.

% \subsection{Contributions and Organization}

In this paper, inspired by \cite{liu2020privacy}, we study Bayesian federated learning protocols based on the digital transmission of gradients from edge devices to the edge server (see Fig. \ref{fig: system model}). Like \cite{liu2021wireless}, which considered analog transmission, we aim at leveraging channel noise for both channel-driven MC sampling and DP.  The main contributions of this paper are as follows.
\begin{itemize}
    \item[$\bullet$] \textbf{Quantized federated LMC (FLMC):} We introduce a quantized federated implementation of LMC based on stochastic quantization, binary transmission,  and  channel-driven sampling;
    \item[$\bullet$] \textbf{Power allocation policy with DP guarantees:} We analyze the DP guarantees of LMC, and we design an approach to determine power control parameter to meet the requirements of both MC sampling and DP;
    \item[$\bullet$] \textbf{Experiments:} We demonstrate an experimental comparison of digital and analog wireless FLMC implementations under DP constraints.
\end{itemize}

The remainder of this paper is organized as follows. Section \ref{sec: system model} formulates the system models and definitions. The privacy anaysis and power control design are presented in Section \ref{sec: power control}. Section \ref{sec: experiments} describes numerical results.

\section{System Model}\label{sec: system model}

As shown in Fig. \ref{fig: system model},  we consider a wireless federated edge learning system comprising an edge server and $K$ edge devices. The devices are connected to the server via a shared wireless channel. Each device $k$ has its own local dataset $\cD_k$, which includes $N_k$  data samples $\cD_k=\{ \bd_{k,n} \}_{n=1}^{N_k}$. The global data set is denoted as  $\cD=\{\cD_k\}_{k=1}^K$. The devices communicate to the server via a NOMA digital channel with BPSK modulation as in \cite{zhu2020one}. Unlike \cite{zhu2020one},  which focuses on conventional frequentist learning, here the goal is to carry out Bayesian learning by approximating the global posterior distribution $p({\bm \theta} | \mathcal{D} )$ at the server. Furthermore, as in \cite{liu2020privacy}, which considers analog transmission, we impose privacy constraints via DP.

\subsection{Federated Langevin Monte Carlo}
The machine learning model adopted by the system is defined by a likelihood function  $p( \bd |{\bm \theta}  )$, as well as by a prior distribution $p(\bm \theta)$. Accordingly, the likelihood of the data at device $k$  is obtained by assuming identical and independent (i.i.d.) observations as
\begin{align}
 p(\cD_k|{\bm \theta}) = \prod_{n=1}^{N_k}  p(\bd_{n,k} |{\bm \theta}).
\end{align}
%, but we will not consider this case explicitly  here. [DZ: Why the difference across devices matters?]} 
The target global posterior is
\begin{align}
 p({\bm \theta}|\cD)  \propto p({\bm \theta}) \prod_{k=1}^K  p(\cD_k | {\bm \theta}), \label{eq: posterior}
\end{align}
which can be expressed in terms of the product $ p({\bm \theta}|\cD)  \propto \prod_{k=1}^K  \tilde{p}({\bm \theta}|\cD_k)$ of the local sub-posteriors at each device $k$
\begin{align}
 &\tilde{p}({\bm \theta}|\cD_k)  \propto p({\bm \theta})^{1/K}  p(\cD_k | {\bm \theta}). \label{eq: local sub-posterior} 
\end{align}
We introduce the local cost function 
\begin{align}\label{eq: local func}
f_k(\bm \theta)=-\log p(\cD_k|{\bm \theta}) -\frac{1}{K}\log p(\bm \theta),
\end{align} 
which accounts for prior and likelihood at device $k$, as well as the global cost function 
\begin{align}\label{eq: gb func}
f(\bm \theta)= \sum_{k=1}^K f_k(\bm \theta). 
\end{align}  

LMC is a gradient-based MCMC sampling scheme.  As such, it aims at producing samples from the global posterior $p({\bm \theta}|\cD)$ in \eqref{eq: posterior} by leveraging information about the gradient of the local cost functions \eqref{eq: local func}. At each $s$-th iteration, LMC produces the next sample $\theta^{[s+1]}$ as
\begin{align} \label{eq: LMC}
\text{(LMC)} \quad {\bm \theta}^{[s+1]}={\bm \theta}^{[s]} -\eta \sum_{k=1}^K \nabla f_k({\bm \theta}^{[s]})  + \sqrt{2\eta} {\bm \xi}^{[s+1]},   
\end{align} 
where $\eta$ is the step size, and $\{{\bm \xi}^{[s]}\}$ is a sequence of i.i.d. random vectors following the Gaussian distribution  $\cN(0, \bI_m)$, which are independent of the initialization ${\bm \theta}^{[0]}\in \mathbb{R}^m$. 

To implement LMC in the described federated setting, 
at each $s$-th communication round, the edge server broadcasts the current sample ${\bm \theta}^{[s]}$ to all edge devices via the downlink channel.  We assume ideal downlink communication.  
%downlink communication is ideal, so that 
%This assumption is practically well justified when the edge sever communicates through a base station with less stringent power constraint than the devices.  
By using the received vector ${\bm \theta}^{[s]}$ and the local dataset $\cD_k$, each device computes the gradient of the local cost function \eqref{eq: local func} as
\begin{align}\label{eq: local gradient}
   \bg_k^{[s]} = -\sum_{n =1}^{N_k}\nabla \log p(\bd_n|{\bm \theta}^{[s]}) -\frac{1}{K} \nabla\log p(\bm \theta^{[s]}). 
\end{align}
While \cite{liu2020privacy} explored the use of analog communication to transmit the local gradients in \eqref{eq: local gradient}, in this work we assume that the devices apply entrywise binary quantization in order to enable BPSK-based transmission. The edge server aggregates the received signals to obtain an approximation of the update term $-\eta \nabla f({\bm \theta}^{[s]})  + \sqrt{2\eta} {\bm \xi}^{[s+1]}$  in \eqref{eq: LMC}. 
%\begin{align} \label{eq: LMC update term}
%-\eta \nabla f({\bm \theta}^{[s]})  + \sqrt{2\eta} {\bm \xi}^{[s+1]} \approx  -\eta \sum_{k=1}^K \tilde{\bg}_k^{[s]}      + \sqrt{2\eta} {\bm \xi}^{[s+1]} . 
%\end{align}
As we will see, and as first proposed in \cite{liu2020privacy},  channel noise can be leveraged to contribute to the additive random term  ${\bm \xi}^{[s+1]}$ in the LMC update \eqref{eq: LMC}, as well as a DP mechanism.    
After $S$ communication rounds, the server obtains a sequence of samples of model parameter vectors $\{{\bm \theta}^{[s]}\}_{s=1}^{S}$. 

\subsection{Communication Model}\label{sec: comm model}
The devices communicate via NOMA on the uplink to the edge server.
%We assume a block flat-fading channel, where the channel coefficients remain constant within a communication block, and they vary in a potentially correlated way over successive blocks.
At any $s$-th communication round, each entry ${\mathrm{g}}_{k,i}^{[s]}$ of the gradient vector ${\bg}_k^{[s]}=[{\mathrm{g}}_{k,1}^{[s]},\cdots, {\mathrm{g}}_{k,m}^{[s]}]^{\sf T}$ is quantized via one-bit stochastic quantization \cite{jin2020stochastic}
\begin{align}
 \tilde{\mathrm{g}}_{k,i}^{[s]}=\begin{cases}
1& \text{with probability $\Phi(\mathrm{g}_{k,i}^{[s]})$}, \\
-1& \text{with probability $1-\Phi(\mathrm{g}_{k,i}^{[s]})$},\label{eq: quantizer}
\end{cases}
\end{align}
where function $\Phi(\cdot)$ returns a probability that increases with the input argument. An example is given by the sigmoid function $\Phi(x)=\sigma(x)=\big(1+\exp(-ax)\big)^{-1}$ for some fixed parameter $a>0$. Each of the quantized gradient parameters $\tilde{\mathrm{g}}_{k,i}^{[s]}$ is modulated into one BPSK symbol.
As a result, a block of $m$ BPSK symbols is produced to communicate  the quantized local gradient vector $\tilde{\bg}_k^{[s]}$ in a communication round. 

Accordingly, at the $s$-th communication round, the received signal at the server is given by the superposition   
 \begin{align} \label{eq: general received signal}
\by^{[s]}= \sum_{k=1}^K\bH_{k}^{[s]}{\bf P}_k^{[s]} \tilde{\bg}_k^{[s]} +\bz^{[s]},
\end{align}
where $\bH_k^{[s]}=\diag[h_{k,1}^{[s]},\cdots, h_{k,m}^{[s]}]$ and ${\bf P}_k^{[s]}=\diag[P_{k,1}^{[s]},\cdots, P_{k,m}^{[s]}]$  are  diagonal matrices collecting respectively the channel gains and power control parameters for $m$ consecutive symbols in a block; while $\bz^{[s]}$ is the channel noise, which is i.i.d. according to distribution $\mathcal{N}(0,{N_0}\bI)$. We assume perfect channel state information (CSI) at all nodes, so that, as we will see, each device can compensate for the phase and amplitude of its own channel.

In the following sections, we will design the power allocation parameters $\{\{P_{k,i}^{[s]}\}_{i=1}^m\}_{k=1}^K$ for each communication round. The transmission of each device is subject to the average per block transmission power constraint:
\begin{equation}\label{eq: power constraint}
{(\text {Power constraint})} \quad  \frac{1}{m}\sum_{i=1}^m \big| P_{k,i}^{[s]} \tilde{\mathrm{g}}_{k,i}^{[s]}\big|^2  \leq P_0, \forall k, s. 
\end{equation}
We define the maximum signal to noise ratio (SNR) as $\SNR_{\sf max}={P_0}/{N_0}$, which is obtained when a device transmits at full power.

\subsection{Differential Privacy} 
We assume an ``honest-but-curious" edge server that may attempt to infer information about local data sets from the received signals $\by^{[s]}$. The privacy constraint is described by the standard $(\epsilon,\delta)$-DP constraint, with some  $\epsilon>0$ and $\delta\!\in\![0,1)$. DP hinges on the divergence between the two distributions $P(\by^{[s]} |\mathcal{D}')$ and $P ( \by^{[s]} |\mathcal{D}'') $ of the signal received when the data sets $\cD'$ and $\cD''$ differ a single data point, i.e., $\|\cD'-\cD'' \|_1=1$. Formally, we have $(\epsilon,\delta)$-DP if the inequality
\begin{equation}\label{eq: def dp}
\max \limits_{\mathcal{D}',\mathcal{D}'':\|\cD'-\cD'' \|_1=1}\big\{\Pr(|\mathcal{L}_{\mathcal{D}',\mathcal{D}''}(\by^{[s]})|\leq \epsilon)\big\}\geq 1-\delta
\end{equation}  
is satisfied, where the DP loss $\mathcal{L}_{\mathcal{D}',\mathcal{D}'' }(\by^{[s]})$ is
\begin{equation}\label{def: privacy loss}
 \mathcal{L}_{\mathcal{D}',\mathcal{D}''}(\by^{[s]} )=\ln \frac{P(\by^{[s]} |\mathcal{D}')}{P ( \by^{[s]} |\mathcal{D}'')}. 
\end{equation} 
The probability in \eqref{eq: def dp} is taken with respect to the distribution $P(\by^{[s]} |\mathcal{D}')$. We note that the DP constraint \eqref{eq: def dp} is applied at each communication round, and that the overall privacy guarantees across iterations can be obtained by using standard composition theorems \cite[Sec. 3.5]{dwork2014algorithmic}. 
To ensure DP requirement as \cite{chen2020understanding,wang2015privacy}, we make the following assumption on the gradients. 

 \begin{assumption}[Bounded Gradients]\label{assumption: BLL}\emph{Each  element of the local gradients is bounded  by some constant $\ell>0$ as 
   \begin{align}\label{eq: BV}
\big| \mathrm{g}_{k,i}^{[s]}  \big| \leq \ell ,\quad  \text{for all }  k, s, i. 
 \end{align}
 }
\end{assumption}
In practice, the condition \eqref{eq: BV} can be met by clipping each entry of the gradient as  $ \min \{1, {\ell}/{| \mathrm{g}_{k,i}^{[s]} |}\} \mathrm{g}_{k,i}^{[s]}$ before quantization \cite{chen2020understanding}.

%\subsection{Assumptions on the Log-Likelihood}
%Finally we make several assumptions on the global cost function in \eqref{eq: gb func} and on its gradient element.
%
%\begin{assumption}[Smoothness]\label{assumption: BLL}\emph{The global function  $f(\bm \theta)$ is smooth with constant $L>0$, which indicates that it is continuously differentiable and the gradient $\nabla f(\bm \theta)$ is Lipschitz continuous with constant $L$, i.e.
%  \begin{align}\label{eq: smoothness}
%  ||\nabla f(\bm \theta)-\nabla f(\bm \theta')||\leq L||\bm \theta-\bm \theta'||
% \end{align}
% for all $\bm \theta, \bm \theta' \in \mathbb{R}^m$. 
% }
%\end{assumption}
%
%\begin{assumption}[Strong Convexity]\label{assumption: BLL}\emph{The global cost function $f(\bm \theta)$ is strongly convex while the following inequality is valid for some constant $\lambda>0$, i.e.
%\begin{align}
%    [\nabla f(\bm \theta)-\nabla f(\bm \theta')]^{\sf T}(\bm \theta-\bm \theta')\geq \lambda||\bm \theta-\bm \theta'||^2
%\end{align}
%for all $\bm \theta, \bm \theta' \in \mathbb{R}^m$. 
% }
%\end{assumption}

\section{Power Control for Quantized Federated Langevin Monte Carlo}\label{sec: power control}

In this section, we first present the transmitter and receiver designs for the proposed quantized federated Langevin Monte Carlo (FLMC), and then analyze its DP properties. Finally, we address the design of power control parameters in \eqref{eq: general received signal}. 
% We note that the design and analysis target for each communication round. For ease of notation, we omit the index of communication round $s$ in the remaining of this section. 
  
 \subsection{Signal Design}

As described in Sec. \ref{sec: comm model}, each device applies stochastic quantization as in \eqref{eq: quantizer}.  Followed by BPSK transmission under the assumption of perfect CSI, we consider channel inversion, whereby the power control matrix in \eqref{eq: general received signal} is selected as $\bP_k^{[s]}=\bA^{[s]} (\bH_k^{[s]})^{-1}$. The diagonal matrix  $\bA^{[s]}=\diag[A_1^{[s]},\cdots, A_m^{[s]}]$ is to be designed with the goal of ensuring that the server can approximate the LMC update \eqref{eq: LMC}, while also guaranteeing the power constraint \eqref{eq: power constraint} and the DP constraint \eqref{eq: def dp}.
%The extension to higher-order modulation is straightforward by simply viewing each modulation symbol as multiple BPSK symbols.  
%We emphasize that, even though we use BPSK modulation in our presentation and the convergence analysis for simplicity, the extension of OBDA to 4-QAM configuration is straightforward by simply viewing each 4-QAM symbol as two orthogonal BPSK symbols. Indeed, we employ 4-QAM modulation for the numerical experiments in Section VI. The long symbol sequence is then divided into blocks, and each block of M symbols is transmitted as a single OFDM symbol with one symbol/parameter over each frequency sub-channel.

The server normalizes the received signal as $(\bA^{[s]})^{-1}\by^{[s]}$ to obtain an estimate of the global gradient. Accordingly, the server approximates the LMC update \eqref{eq: LMC} as 
 \begin{align} \label{eq: oma update}
   {\bm \theta}^{[s+1]}=  {\bm \theta}^{[s]}- \eta \l[ \sum_{k=1}^K \tilde{\bg}_k^{[s]}+ \big(\bA^{[s]}\big)^{-1} \bz^{[s]} \r]. 
 \end{align}  
 
\subsection{Privacy Analysis}
We now consider the DP constraint \eqref{eq: def dp} for any device $k$. To this end, we fix the quantized gradients $\{\tilde{\bg}_j\}_{j \neq k}$ of the other devices, and consider neighboring data sets $\cD_k'$ and $\cD_k''$ for device $k$ that differ only by one sample, i.e., $\|\cD_k'-\cD_k''\|_1=1$. As the DP constraint \eqref{eq: def dp} is applied to every iteration, we omit the index of the communication round $s$ for ease of notation. Then, the privacy loss \eqref{def: privacy loss} for device $k$ can be written as 
\begin{align}
&\mathcal{L}_{\mathcal{D}',\mathcal{D}''}( \by )
= \ln\frac{\prod_{i=1}^m P(A_i \tilde{\mathrm{g}}'_{k,i} +A_i \sum_{q\neq k}\tilde{\mathrm{g}}_{q,i} +z_i \big|\{\tilde{\mathrm{g}}_{q,i}  \}_{q\neq k}, \cD_k' )  }{\prod_{i=1}^m  P(A_i \tilde{\mathrm{g}}''_{k,i} +A_i \sum_{q\neq k}\tilde{\mathrm{g}}_{q,i} +z_i \big|\{\tilde{\mathrm{g}}_{q,i}  \}_{q\neq k}, \cD_k'')  } \nn  \\ 
&=\sum_{i=1}^m\ln\frac{\l[ \Phi(\mathrm{g}_{k,i}') \exp\l( \frac{2  (z_i-A_i \sum_{q\neq k } \tilde{\mathrm{g}}_{q,i})}{N_0/A_i}\r)+\l(1-\Phi(\mathrm{g}_{k,i}')\r)\r]}{\l[\Phi(\mathrm{g}_{k,i}'')\exp\l( \frac{2  (z_i-A_i \sum_{q\neq k } \tilde{\mathrm{g}}_{q,i})}{N_0/A_i}\r)+\l(1-\Phi(\mathrm{g}_{k,i}'')\r)\r]},\label{eq: digital privacy format}
\end{align}
where, with some abuse of notation, $P(X|Y)$ represents the distribution of random variable $X$ evaluated at $X$ when conditioned on the value $Y$ of random variable $Y$; 
the last step uses the fact that the distributions in \eqref{eq: digital privacy format} are mixture of Gaussians; and we have $z_i\sim\mathcal{N}(0,N_0)$. To attain the maximum DP loss in \eqref{eq: digital privacy format}, we consider the worst-case choice of data sets $\mathcal{D}'$ and $\mathcal{D}''$. To this end, without loss of generality, we set $\Phi(\mathrm{g}_{k,i}')=\Phi(\ell)$ and $\Phi(\mathrm{g}_{k,i}'')=\Phi(-\ell)$ by Assumption~\ref{assumption: BLL}. Furthermore, the value of the sum $\sum_{j\neq k}\tilde{\mathrm{g}}_{q,i}$ is within the range of $[-(K-1),(K-1)]$, and hence have the following inequality 
\begin{align}
&|\mathcal{L}_{\mathcal{D}',\mathcal{D}''}(\by)|\nn \\ \leq &\max\Bigg\{  \Bigg|\sum_{i=1}^m\ln\frac{\l[ \Phi(\ell) \exp\l( \frac{2  (z_i+A_i (K-1))}{N_0/A_i}\r)+\l(1-\Phi(\ell)\r)\r]}{\l[\Phi(-\ell)\exp\l( \frac{2  (z_i+A_i (K-1))}{N_0/A_i}\r)+\l(1-\Phi(-\ell)\r)\r]}\Bigg|, \nn\\
& \Bigg| \sum_{i=1}^m\ln\frac{\l[ \Phi(\ell) \exp\l( \frac{2  (z_i-A_i (K-1))}{N_0/A_i}\r)+\l(1-\Phi(\ell)\r)\r]}{\l[\Phi(-\ell)\exp\l( \frac{2  (z_i-A_i (K-1))}{N_0/A_i}\r)+\l(1-\Phi(-\ell)\r)\r]} \Bigg| \Bigg\} \nn \\
&\triangleq \mathcal{L}^*(\bz),
\label{eq: worst case privacy loss}
\end{align}
where $\bz \sim \mathcal{N}(0,\bI_m)$. We can now use \eqref{eq: worst case privacy loss} to evaluate numerically a bound on left-hand side of \eqref{eq: def dp} as $\Pr(|\mathcal{L}^*(\bz)|\leq\epsilon)\geq1-\delta$ with $\bz \sim \mathcal{N}(0,\bI_m)$.

To compare with analog FLMC in \cite{liu2020privacy},   we reproduce the privacy loss in \cite{liu2020privacy} as 
\begin{align}
    \mathcal{L}_{\mathcal{D}',\mathcal{D}''}(\by)=\sum_{i=1}^m\frac{2z_iA_i\Delta_{k,i}+(A_i\Delta_{k,i})^2}{2N_0}, \label{eq: analog dp constraint}
\end{align}
where $z_i \sim\mathcal{N}(0,N_0)$, and $\Delta_{k,i}=|\mathrm{g}'_{k,i}-\mathrm{g}'_{k,i}|$, and we have $\Delta_{k,i}\leq 2\ell$. To gain some insight about the comparison between \eqref{eq: worst case privacy loss} and \eqref{eq: analog dp constraint}, consider the high-SNR regime in which the power of channel noise $N_0$ approaches $0$. In this case, the privacy loss \eqref{eq: analog dp constraint} in the analog scheme goes to infinity, and hence no $(\epsilon,\delta)$-DP level with $\delta<1$ is possible. This is in sharp contrast with the digital scheme, for which the privacy loss \eqref{eq: worst case privacy loss} is upper bounded by $m \ln  \Phi(\ell) - m \ln \Phi(-\ell)$. This discussion illustrates the potential advantages of the digital scheme in the presence of privacy constraints in the high-SNR regime. 

%Stochastic quantization introduces additional protection in DP. For example, if the variance of the channel noise $N_0$ approximates to zero which leads to high SNR, the DP constraint on digital FLMC can be treated as $\Pr(m\ln \frac{\Phi(\ell)}{\Phi(-\ell)}\leq\epsilon)\geq1-\delta$ and we can achieve $(0,0)$-DP with equal probability quantization. But with this assumption, the privacy loss for analog FLMC in \eqref{eq: analog dp constraint} approximates to infinity, which in turn incurs infinity privacy level $\epsilon$ and $\delta=1$. This property is in sharp contrast with high SNR analog transmission where zero privacy loss is unreachable.

\subsection{Power Control}
The design of power  control  parameters in the power gain matrix $\bA^{[s]}$ must comply with the power constraints, the LMC noise requirements, and the DP constraints. 

For the power constraint \eqref{eq: power constraint}, plugging in the choice $\bP_k^{[s]}=\bA^{[s]} (\bH_k^{[s]})^{-1}$ yields the inequalities
\begin{align}
\frac{1}{m}\sum_{i=1}^m \l(\frac{ A_i^{[s]}}{ h_{k,i}^{[s]}} \r)^2    \leq  {P_0}, \  \forall k, s.\label{eq: numerical power constraint}
\end{align}

Furthermore, in order to guarantee that the noise powers $N_0\eta^2 (A_i^{[s]})^{-2}$ in the update \eqref{eq: oma update} are no smaller than the power $2\eta$ required by the LMC update \eqref{eq: LMC} we impose the LMC noise requirement (see also \cite{liu2021wireless}) 
\begin{align}\label{eq: numerical LMC noise requirement}
A_i^{[s]} \leq \sqrt{\frac{\eta N_0}{2}}, \ \forall i, s. 
\end{align}
Finally, to impose the DP constraint, given the desired level of privacy loss $\epsilon$,  we numerically estimate the probability $\delta$ in \eqref{eq: def dp} as a function of power gain parameters $A_i^{[s]}$   by drawing samples from the noise $\bz^{[s]} \sim \cN (0, N_0 \bI)$.

%
%
%The choice of the power gain parameter $A_i^{[s]}$ should be set as the minimum value of power constraint \eqref{eq: numerical power constraint}, LMC noise requirement \eqref{eq: numerical LMC noise requirement} and the DP constraint.

\section{Numerical Results}\label{sec: experiments}
In this section, we evaluate the performance of the proposed quantized FLMC, and compare it with the analog transmission scheme introduced in \cite{liu2021wireless}. Throughout this section, we assume the channel coefficients to be constant within a communication block, and homogeneous across the devices, i.e., $h_{k,i}^{[s]}=h^{[s]}$ for all devices $k=1,\cdots, K$ and all elements  $i=1,\cdots, m$. Under this assumption, the power gains for quantized FLMC are obtained via a numerical search to maximize the value of $A_i^{[s]}$ under the three constraints reviewed in the previous sections. In a similar manner, for {\bf analog FLMC}, we have \cite{liu2020privacy}
\begin{align}
A_i^{[s]} = \min \l\{ \frac{|h^{[s]}|\sqrt{P_0}}{\ell}, \sqrt{\frac{\eta N_0}{2}}, \sqrt{\frac{N_0\cT^{-1}(1-\delta)}{2m\ell^2}}\r\}, \ \forall k,\ s, \label{eq: analog constraints}
\end{align}
where the last term is the inverse function of $\cT(x)$  defined by the  error function ${\displaystyle \operatorname {erf} (x)={\frac {2}{\sqrt {\pi }}}\int _{0}^{x}e^{-t^{2}}\,dt}$ as
\begin{align}
    \cT(x)={\rm erf}\l(\frac{\epsilon-x}{2\sqrt{x}}\r)-{\rm erf}\l(\frac{-\epsilon-x}{2\sqrt{x}}\r),
\end{align}
which is obtained by plugging \eqref{eq: analog dp constraint} into \eqref{eq: def dp}, and leveraging the tail probability of Gaussian distribution. We also consider benchmark schemes without DP constraint.

As for the learning model, as in \cite{liu2021wireless}, we consider a Gaussian linear regression with likelihood
 \begin{align}\label{eq: gl model}
p(v_n|{\bm \theta}, \bu_n) =  \frac{1}{\sqrt{2\pi}}e^{-\frac{1}{2} (v_n-{\bm \theta}^{\sf T}\bu_n)^2}, 
\end{align}
and the prior $p({\bm \theta})$ is assumed to follow Gaussian distribution $\cN(0, \bI_m)$. Therefore, the posterior $p({\bm \theta}|\cD)$ is the Gaussian $\cN\big((\bU\bU^{\sf T}+\bI)^{-1}\bU\bv,(\bU\bU^{\sf T}+\bI)^{-1}\big)$,  where $\bU=[\bu_1,\cdots,\bu_N]$ is the data matrix and $\bv=[v_1, \cdots, v_N]^{\sf T}$ is the label vector. We use synthetic dataset $\{\bd_n=(\mathbf{u}_n,v_n)\}_{n=1}^{N}$ with $N=1200$ following the learning model in \eqref{eq: gl model}, with input $\mathbf{u}_n$ drawn i.i.d from $\mathcal{N}(0,\mathbf{I}_m)$ where $m=5$.  The ground-truth model parameter is $\bm\theta^*=[0.418, -0.289, 0.3982, 0.8231, 0.5251]^{\sf T}$. 
Unless stated otherwise, the data set is evenly distributed to $K=20$ devices; the constant channel $h^{[s]}$ is set to $0.04$ for all communication rounds; the power of channel noise is set to $N_0=1$; the bound of gradient element is set to  $\ell=30$; learning rate is set to $\eta=1.28\times 10^{-4}$ for analog FLMC and $\eta=8.28 \times 10^{-3}$ for digital FLMC, which are tuned by using the smoothness and strongly convexity parameters (see \cite{liu2021wireless}). We consider a sigmoid  function for  quantization probability in \eqref{eq: quantizer} as $\Phi(x)=[1+\exp(- ax)]^{-1}$, and set $a=0.05$ by default.

The total number of communication rounds is chosen as $S=300$, which are comprised of $S_b=200$ samples for the burn-in period, and the following $S_u=S-S_b=100$ samples for evaluation. The quality of the samples is measured by mean squared error (MSE) 
 \begin{align}
    \text{MSE}=\frac{1}{S_u}\sum_{s=S_b+1}^{S_b+S_u}\|\bm \theta^{[s]}-\bm \mu\|^2,
    \label{eq:error}
\end{align}
where $\boldsymbol{\mu}$ is the mean of the ground-truth posterior distribution. All the results are averaged over 1000 experiments. 

\begin{figure}[t]

%\begin{minipage}[b]{1.0\linewidth}
  \centering
  \centerline{\includegraphics[scale=0.58]{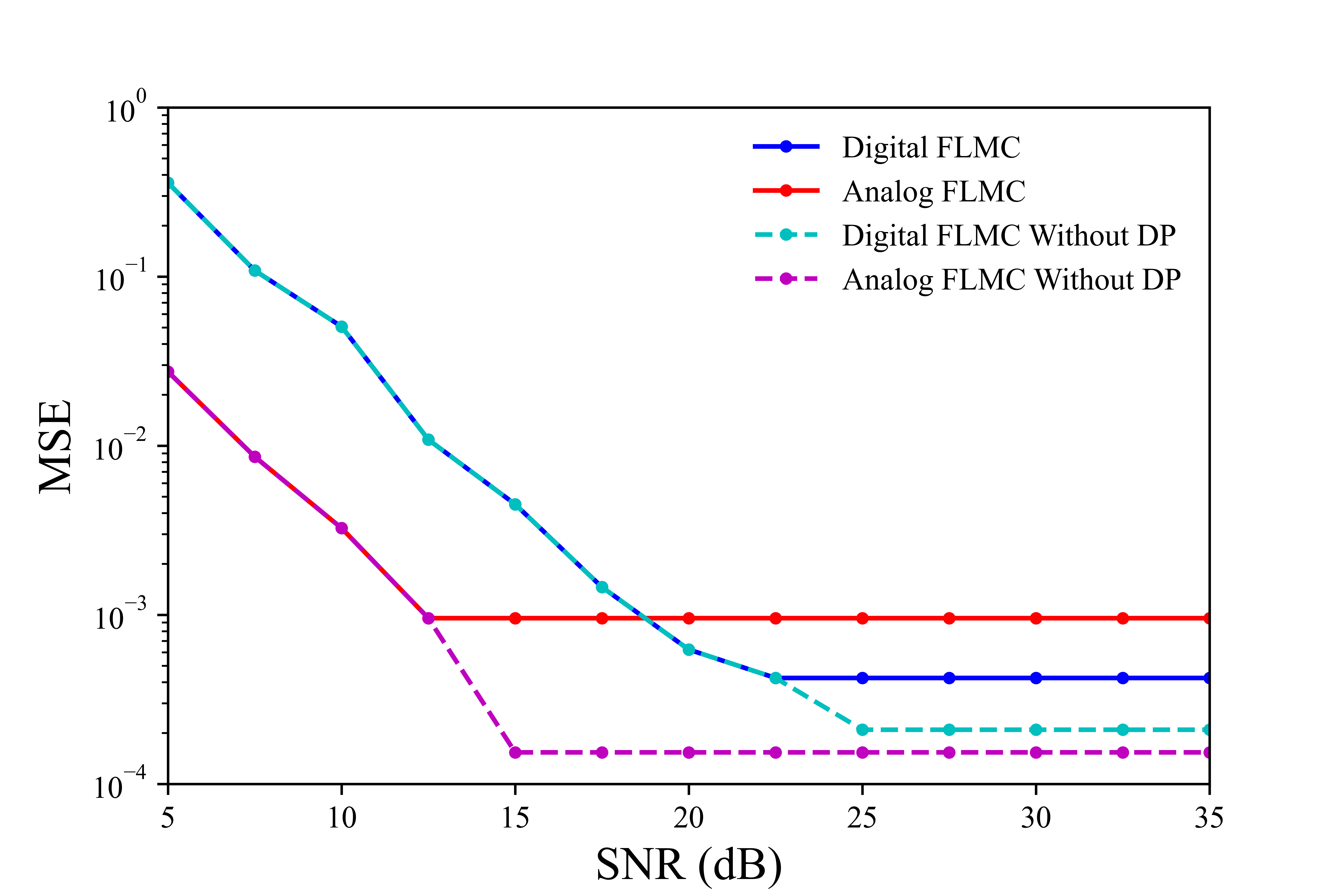}}
  \vspace{-0.2cm}
  \caption{MSE as a function of SNR ($\epsilon=5, \delta=0.01$).}\label{fig: snr_impact}
%
%\vspace{-0.2cm}
\end{figure}

We first investigate the impact of SNR in Fig. \ref{fig: snr_impact} on the performance of digital and analog FLMC schemes.  In this experiment, we set the DP level as $\epsilon=5$ and $\delta=0.01$. Confirming the discussion in the previous section, in the high-SNR regime, digital FLMC is seen to outperform analog FLMC, since the latter one must back off the transmitted power in order to meet the DP constraint. In contrast, SNR lower than $17.5$ dB, analog FLMC is preferable.

% performance of digital transmission is limited by power constraint until $\SNR_{\sf max}=23$ dB, after which it becomes DP limited. However, analog FLMC reaches to DP-limited regime when $\SNR_{\sf max}=10$ dB. In the DP-limited regime, digital FLMC attains better performance than analog FLMC.

\begin{figure}[t]

%\begin{minipage}[b]{1.0\linewidth}
  \centering
  \centerline{\includegraphics[scale=0.58]{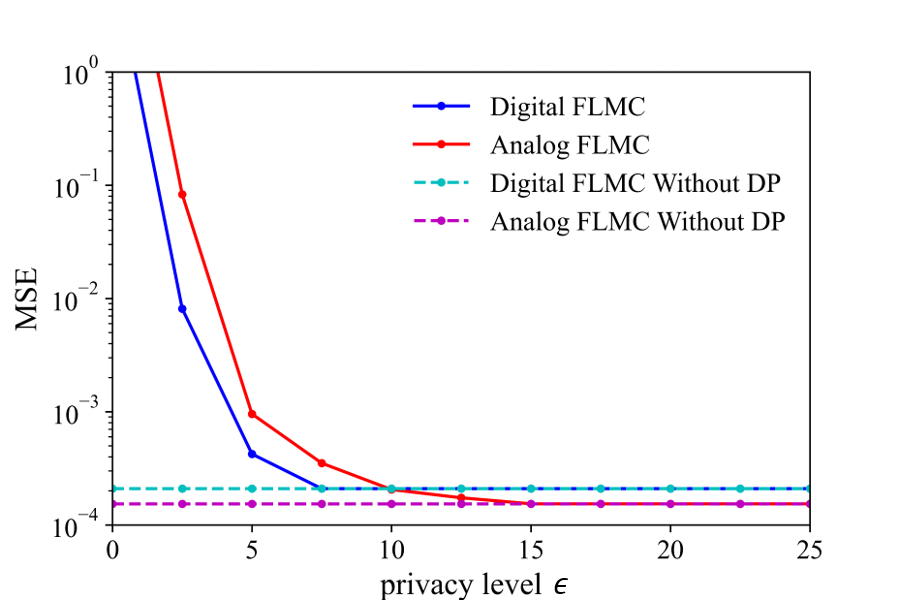}}
  \vspace{-0.2cm}
  \caption{MSE as a function of privacy level $\epsilon$ ($\SNR_{\sf max}=25$ dB, $\delta=0.01$).}\label{fig: epsilon_impact}
\vspace{-0.5cm}
\end{figure}

We now further investigate the impact of the privacy level on the digital and analog FLMC schemes in Fig. \ref{fig: epsilon_impact}.  In this experiment, we set $\SNR_{\sf max}=25$ dB. The error of all schemes is seen to decrease by relaxing the DP constraint, until $\epsilon=7.5$ for the digital scheme and $\epsilon=15$ for the analog scheme. Relaxing the DP constraint cannot reduce the error, as the performance becomes limited by the transmitted power constraint or by LMC noise requirement. The digital FLMC scheme outperforms analog FLMC under a stricter DP requirement, i.e., when $\epsilon\leq 7.5$. This provides further validation of the advantage of the digital scheme when the SNR is large enough. 
 
\begin{figure}[t]

%\begin{minipage}[b]{1.0\linewidth}
  \centering
  \centerline{\includegraphics[scale=0.58]{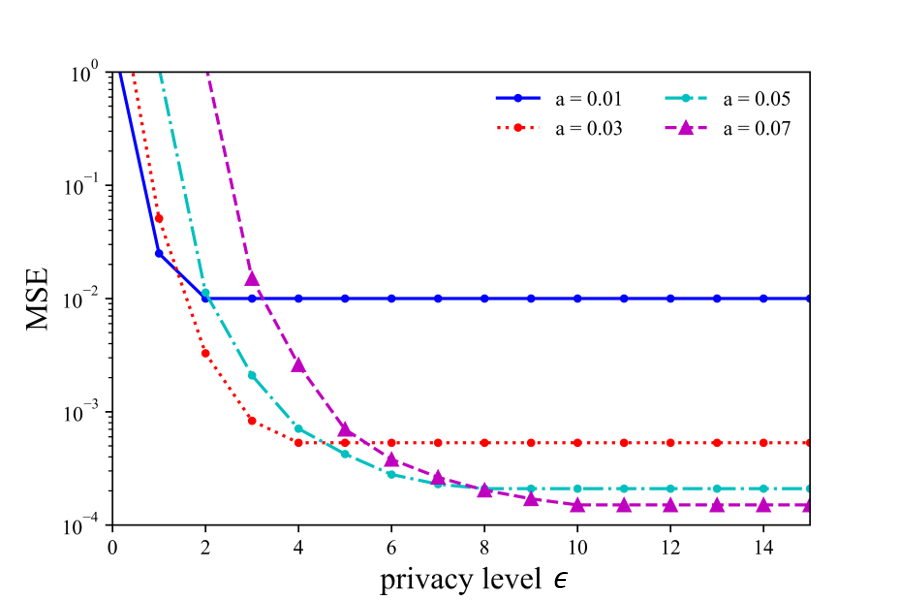}}
  \vspace{-0.2cm}
  \caption{MSE as a function of privacy level $\epsilon$ for different parameter of the stochastic binary quantization $a$ ($\SNR_{\sf max}=25$ dB, $\delta=0.01$).}\label{fig: bandwidth_impact}
%
%\vspace{-0.2cm}
\end{figure}

Finally, in Fig. \ref{fig: bandwidth_impact}, we study the impact of varying the parameter $a$ of the quantization probability function $\Phi(x)=[1+\exp(- ax)]^{-1}$. Note that a small $a$ implies a more noisy quantizer. In this experiment, we also set $\SNR_{\sf max}=25$ dB. Under strict DP requirement $\epsilon<2$, the quantizer with the small value $a=0.01$ outperforms other choices, since the higher level of randomness is applied to meet the DP constraint. Conversely, by relaxing the DP requirement, quantizer with larger value of $a$ become advantageous.

%With more relaxing DP constraint after $\epsilon=8$, quantizer with $a=0.07$ bandwidth setting obtains best performance since it leads to more deterministic quantization while others have less contributions to the LMC update. On the other hand, with stricter DP constraint $\epsilon<2$, quantizer with bandwidth $a=0.01$ outperforms other settings due to the fact that privacy loss is lower under more stochastic quantization.
\vspace{-0.2cm}

%\section{Conclusions}
%In this paper, we have proposed a Bayesian federated learning (FL) protocol that implements one-bit federated LMC via shared wireless transmission from edge devices to edge server.

\bibliographystyle{ieeetr}
\bibliography{DP_Fed_ref}
\end{document}